\documentclass[manuscript,showpacs]{revtex4}
\usepackage{graphicx}
\usepackage{amsmath}
\usepackage{amssymb}
\begin{document}

\title{Entanglement enhanced atomic gyroscope}
\author{J.J. Cooper, D.W. Hallwood, and J. A. Dunningham}
\affiliation{School of Physics and Astronomy, University of Leeds, Leeds LS2 9JT, United Kingdom}

%================================================
\begin{abstract}
The advent of increasingly precise gyroscopes has played a key role in the technological development of navigation systems. Ring-laser and fibre-optic gyroscopes, for example, are widely used in modern inertial guidance systems and rely on the interference of unentangled photons to measure mechanical rotation.  The sensitivity of these devices scales with the number of particles used as $1/\sqrt{N}$.  Here we demonstrate how, by using sources of entangled particles, it is possible to do better and even achieve the ultimate limit allowed by quantum mechanics where the precision scales as $1/N$. We propose a gyroscope scheme that uses ultra-cold atoms trapped in an optical ring potential.
\end{abstract}
%================================================

\maketitle

\section{Introduction}
Optical interferometers have revolutionised the field of metrology,  enabling path length differences to be measured, for the first time,  to less than the wavelength of the light being used. As well as their more familiar linear versions, interferometers can also be used in ring geometries to make accurate measurements of angular momentum. These interferometric gyroscopes surpass the precision of their mechanical counterparts and form a key component of many modern navigation systems.
They work by exploiting the different path lengths experienced by light as it propagates in opposite directions around a rotating ring. 
For instance, in the Sagnac geometry \cite{Sagnac1913} photons are put into a superposition of travelling in opposite directions around a ring and, when rotated, the two directions acquire different phases. This phase difference is directly related to the rate of rotation and can be measured by  recombining the two components at a beam splitter and recording the intensity at each of the outputs. 

Such schemes use streams of photons that are independent of one another, i.e. not entangled. In this case the measurement accuracy is fundamentally limited by the jitter in the recorded intensities due to the fact that photons come in discrete packages. This is known as the shot-noise limit and restricts the measurement to a precision that scales inversely with the square root of the total number of photons.  One possible way to beat this precision limit is to use entangled particles \cite{Yurke86}.  In fact it has long been known that the precision of optical interferometers is improved with the use of squeezed or entangled states of light \cite{Caves1981, Lee2002, Giovannetti2004}.  In principle, this should enable us to reach the Heisenberg limit whereby the precision scales inversely with the total number of particles.

The precision of gyroscopic devices can be improved further still with the use of entangled atomic states due to their mass enhancement factor over equivalent photonic devices \cite{Dowling1998}.  The use of entangled atomic states to make precision measurements of rotations was first proposed in \cite{Dowling1998}.  Since then, and with the experimental realization of Bose-Einstein condensates, the use of entangled atoms for precision measurements has been widely researched.  As in the optical case, of key importance to the ultimate precision afforded by an atomic device is the input state used.  The use of number-squeezed atomic states will allow for Heisenberg limited precisions and as such there has been much research into the generation and uses of these squeezed states \cite{Orzel2001, Li2007, Esteve2008,  Haine2009}.  Several proposals have already been made to use these squeezed, and other entangled atomic states, such as maximally entangled `NOON' states, to make general phase measurements with sub-shot noise sensitivities \cite{Bouyer1997, Dunningham04, Pezze2005, Pezze2006, Dunningham01a, Pezze2007, Pezze2009}.  
It has also been shown that uncorrelated atoms can also achieve sub shot-noise sensitivities of rotational phase shifts \cite{Search2009} by using a chain of matter wave interferometers, or a chain of gyroscopes.  

Here we propose a gyroscope scheme that uses squeezed and entangled atomic inputs to push the sensitivity of rotational phase measurements below the shot noise limit.  We extend the investigation of optimal input states to an experimentally accessible atomic gyroscope scheme capable of measuring small rotations.  It works by trapping ultra-cold atoms in a one-dimensional optical lattice in a ring geometry and carefully evolving the trapping potential.  We investigate the precision achieved with different inputs and show that although a so-called NOON state produces the best precision, another state (sometimes referred to as a `bat' state) that is created by passing a number-squeezed state through a beam splitter, is far more robust and might therefore be a preferred candidate.

It should be noted that it is possible to beat the Heisenberg precision scaling of $1/N$ in some measurement schemes.  In fact precision scalings of $N^{-3/2}$ can be achieved even when the initial state is unentangled in a few specific metrology protocols \cite{Boixo2008, Boixo2009a}.  This is achieved through a nonlinear coupling between the quantum probe and the parameter to be measured.  Therefore, BECs with their particle interactions, may naturally lend themselves to this.  However, the challenge would be to find ways of coupling the angular momentum we wish to measure to the scattering length of the atoms.  While this may provide an interesting future direction to this work, in light of this difficulty, we concentrate for now on attaining the Heisenberg limit through optimizing the input state.

\section{The system}
Our system consists of a collection of ultra-cold atoms trapped by the dipole force in an optical lattice loop of three sites. Rings with this geometry have already been experimentally demonstrated \cite{Boyer2006, Henderson2009}. For a sufficiently cold system, we need only consider a single level in each site and so can describe the system using the Bose-Hubbard Hamiltonian,
\begin{equation} \label{ham1}
\frac{H}{\hbar}=\sum_{j=0}^{2}\epsilon_{j}a_{j}^{\dag}a_{j} -\sum_{j=0}^{2}J_{j}\left(a_{j}^{\dag}a_{j+1} + a_{j+1}^{\dag}a_{j} \right)+\sum_{j=0}^{2}V_{j}a_{j}^{\dag}{}^{2}a_{j}^{2},
\end{equation} 
where $a_{j}$ is the annihilation operator for an atom at site $j$ and $J_j$ is the coupling strength between sites $j$ and $j+1$.  The ring geometry means that $a_{j}=a_{j+3}$.  The parameter $V_{j}$ is the strength of the interaction between atoms on site $j$ and $\epsilon_{j}$ accounts for the energy offset of site $j$.  In general, we take the zero point energy to be the same for each site and so set $\epsilon_j=0$.  For the purposes of this work it will be convenient to describe the system in terms of the quasi-momentum (or 
flow) basis which is related to the site basis by,
\begin{equation}
\alpha_k = \frac{1}{\sqrt{3}}\sum_{j=0}^{2} e^{i2\pi jk/3}a_{j},
\label{fourier}
\end{equation} where $\alpha_k$ corresponds to the annihilation of an atom with $k$ quanta of flow.  We shall use positive subscripts to refer to clockwise flow and negative subscripts to refer to anti-clockwise flow. 

\section{Scheme 1: Uncorrelated particles}
The first scheme we present uses unentangled atoms to achieve shot noise limited precisions.  To begin with, the potential barriers between the sites are high and $N$ atoms are contained within one site, say site zero.  The initial state of the system is therefore $|\psi\rangle_{U0}=|N,0,0\rangle$ where the terms in the ket represent the number of atoms in sites zero, one and two respectively.

The first step is to rapidly reduce the potential barrier between just two sites, we choose sites zero and one, in such a way that the two sites remain separate but there is strong coupling between them.  This must be done rapidly with respect to the tunneling time, but slowly with respect to the energies associated with excited states in order to ensure the system remains in the ground state.  This separation of timescales has already been demonstrated experimentally \cite{Greiner2002a}.  In this regime the coupling between the two sites is much larger than their on site interactions and the Hamiltonian describing the two sites is,
\begin{equation}
\frac{H_{2J}}{\hbar}=-J(a_0^\dag a_1+a_1^\dag a_0) .
\label{H2J}
\end{equation}
Importantly, the remaining two barriers are high ($V \gg J$) and so prevent tunneling between sites one and two and sites two and zero.  

The system is left to evolve for time $t=\pi/4J$ whilst this barrier is low.  This is equivalent to applying a two port 50:50 beam splitter to our initial state (as shown in \cite{Dunningham01a}) and so transforms $|\psi\rangle_{U0}$ to,
\begin{equation}
|\psi\rangle_{U1} = \frac{1}{\sqrt{2^{N}N!}}(a_0^\dag + ia_1^\dag)^N|0,0,0\rangle .
\label{psi1s1}
\end{equation}
Each individual atom is now equally likely to be on site zero or one.  In other words we have $N$ single-particle superpositions on the two sites.

The next step is to apply a three port beam splitter, or tritter.  This splitting procedure is described in detail in \cite{Cooper09}.  Essentially to achieve a tritter in this system we immediately lower the two remaining potential barriers, on the same timescale as before, and allow the system to evolve for a further $t=2\pi/9J$.  This tritter operation is given by,
\begin{equation}
S_3=
\frac{1}{\sqrt{3}}
\begin{pmatrix}
1 & e^{i2\pi /3} & e^{i2\pi /3} \\
e^{i2\pi /3} & 1 & e^{i2\pi /3} \\
e^{i2\pi /3} & e^{i2\pi /3} & 1
\end{pmatrix}
\label{tritter}
\end{equation}
from which we can see $|\psi\rangle_{U1}$ is transformed to,
\begin{equation}
|\psi\rangle_{U2} = \frac{1}{\sqrt{2^{N}3^{N}N!}}\left( (a_0^\dag + e^{i2\pi /3}a_1^\dag + e^{i2\pi /3}a_2^\dag)+i(e^{i2\pi /3}a_0^\dag + a_1^\dag + e^{i2\pi /3}a_2^\dag) \right)^{N}|0,0,0\rangle .
\end{equation}
At this point we rapidly raise the potential barriers, `freezing' the atoms in the lattice sites.  Comparing $|\psi\rangle_{U2}$ with equation (\ref{fourier}) we see that applying a $2\pi/3$ phase to site two results in a superposition of the $\alpha_{-1}$ and $\alpha_{1}$ flow states.  This phase is achieved by applying an energy offset, $\epsilon_2$, to site two, whilst the barriers are high, for time $t_{\epsilon}=4\pi/3\epsilon_2$.  Offset application times of $500$ns have been demonstrated experimentally \cite{Denschlag2000} and it is this time we shall use in section \ref{Interactions} when we assess the impact of non-zero interactions.  

We then immediately lower the barriers again so the atoms can flow around the loop.  The resulting superposition can be written as,
\begin{equation}
|\psi\rangle_{U3} = \frac{1}{\sqrt{2^{N}N!}}(\alpha_{-1}^{\dag} + i\alpha_{1}^{\dag})^{N}|0,0,0\rangle 
\end{equation}
where the terms in the ket now represent the number of atoms in each of the possible flow states, $\alpha_{-1}$, $\alpha_0$ and $\alpha_1$ respectively.  This is now a $N$ single particle flow superposition.

At this point the $\alpha_{-1}$ and $\alpha_{1}$ states are degenerate, so $|\psi\rangle_{U3}$ does not evolve.  However, we now apply the rotation we wish to measure, $\omega$, to the ring which causes a phase, $\theta$, to be applied around it.  The energies of the two flow states now change according to the Hamiltonian, 
\begin{equation}
\frac{H_k}{\hbar} = -2J\sum_{k=-1}^1 \cos\left(\theta/3 - 2\pi k/3\right)\alpha_k^{\dag}\alpha_k .
\end{equation}
After a time $t_{\omega}$, and ignoring global phases, the state has evolved to
\begin{equation}
|\psi\rangle_{U4} = \frac{1}{\sqrt{2^{N}N!}} \left(e^{i2Jt_{\omega}\cos(\theta/3+2\pi/3)}(\alpha_{-1}^{\dag})+ie^{i2Jt_{\omega}\cos(\theta/3-2\pi/3)}(\alpha_{1})^{\dag} \right)^{N}|000\rangle
\end{equation}
and so a phase difference of $\phi = 2\sqrt{3}Jt_{\omega}\sin(\theta/3)$ is established between the two flows.

We now wish to read-out this phase difference from which we can directly determine $\omega$ since $\omega = h\theta/({L^2 m})$ where $m$ is the mass of the atom and $L$ is the circumference of the ring.  The read-out procedure involves sequentially undoing all the operations performed prior to the phase shift.  This is analogous to standard Mach-Zehnder interferometry where an (inverse) beam splitter is placed after the phase shift to undo the initial beam splitting operation.  

The undoing process begins with the application of a $-2\pi/3$ phase to site two giving,
\begin{equation}
|\psi\rangle_{U5} = \frac{1}{\sqrt{2^{N}3^{N}N!}}\left((a_0^\dag + e^{i2\pi /3}a_1^\dag + e^{i2\pi /3}a_2^\dag)+ie^{i\phi}(e^{i2\pi /3}a_0^\dag + a_1^\dag + e^{i2\pi /3}a_2^\dag) \right)^N|0,0,0\rangle
\end{equation}
where the terms in the kets now, once again, represent the number of atoms in sites zero, one and two.  Next we undo the tritter by applying an inverse tritter, $S_{3}^{-1}=S_{3}^\dag$.    The inverse tritter operation is discussed in detail in \cite{Cooper09} but essentially is achieved by lowering all three barriers and allowing the system to evolve for time $t=4\pi/9J$ (i.e. twice as long as for a tritter) giving,
\begin{equation}
|\psi\rangle_{U6} = \frac{1}{\sqrt{2^{N}N!}} \left(a_0^\dag + ie^{i\phi}a_1^\dag\right)^N |0,0,0\rangle
\label{output}
\end{equation}
which is equivalent to $|\psi\rangle_{U1}$ but with a phase difference, $\phi$.  

Finally we apply an inverse two port 50:50 beam splitter.  This is achieved in just the same way as the two port beam splitting operation described above, but with a hold time of $t=3\pi/4J$ rather than  $t=\pi/4J$.  The resulting state is,
\begin{equation}
|\psi\rangle_{U7} = \frac{1}{\sqrt{N!}} \left( \cos \left(\frac{\phi}{2}\right)(a_0^\dag) - \sin \left(\frac{\phi}{2}\right)(a_1^\dag)  \right)^N |0,0,0\rangle
\end{equation}
meaning the probabilities of detecting each atom at site zero and site one are,
\begin{eqnarray}
P_0 = \cos^2\left(\frac{\phi}{2} \right) \\ \nonumber
P_1 = \sin^2\left(\frac{\phi}{2} \right) .
\label{probS3}
\end{eqnarray}
Since the atoms are independent the total number detected in the two sites is given by a binomial distribution.  The mean number of atoms detected at site zero is therefore $\langle n_0 \rangle = N\cos^2(\phi/2)$ and at site one it is $\langle n_1 \rangle = N\sin^2(\phi/2)$.  By counting the number of atoms detected at each site we can determine $\phi$, and hence $\omega$, just as in a typical Mach-Zehnder interferometer.

The precision with which this scheme enables us to measure $\omega$ can be found by calculating the quantum Fisher information, $F_Q$. This is a tool for evaluating the precision limits of quantum measurements and is independent of the measurement procedure.  For a pure state $|\Psi(\phi)\rangle$ it is given by~\cite{Braunstein1994},
\begin{equation}
F_Q=4\left[ \langle\Psi'(\phi)|\Psi'(\phi)\rangle - \left|  \langle\Psi'(\phi)|\Psi(\phi)\rangle \right|^2 \right]  .
\label{Fq}
\end{equation}
We convert this into an uncertainty in $\phi$ using the Cramer-Rao lower bound \cite{Rao1945, Cramer1946, Helstrom1976}, 
\begin{equation}
\Delta \phi \geq 1/ \sqrt{F_Q}.
\label{Cramer}
\end{equation} 
Using $|\psi\rangle_{U4}$ we find the maximum resolution scaling of $\phi$ with $N$ is $N^{-1/2}$, or equivalently $\Delta\theta \sim \sqrt{3}/(2Jt_{\omega}\cos(\theta/3)\sqrt{N})$.  This translates to an uncertainty in $\omega$ of,
\begin{equation}
 \Delta \omega \sim \left(\frac{h}{L^2 m}\right)\frac{\sqrt{3}}{2Jt_{\omega}\sqrt{N}}
 \end{equation}
 where we have made the approximation that $\theta/3 \ll 1$.  This has the well-known $1/\sqrt{N}$ scaling that is a signature of the shot-noise limit.  

To summarize, shot-noise limited measurements of rotations are made as follows:\\
1.  Apply a two port 50:50 beam splitter to the first two modes of the state $|N,0,0\rangle$.\\
2.  Perform a three port beam splitter (tritter) operation to the state.\\
3.  Apply a $2\pi/3$ phase to site two.\\
4.  Leave the system to evolve for time $t_{\omega}$ under the rotation, $\omega$.\\
5.  Apply a $-2\pi/3$ phase to site two.\\
6.  Perform an inverse tritter operation on the state.\\
7.  Apply an inverse two port 50:50 beam splitter to the first two modes.\\
8.  Count the number of atoms in each site.

We will now show how our scheme can be modified to create entangled states and will investigate the effect of this entanglement on the precision scaling of our rotation measurements.

%===================================================================================================================================
\section{Scheme 2: The bat state}\label{bat}

We begin with $N/2$ atoms on site zero and on site one, i.e. $|\psi\rangle_{B0} = |N/2,N/2,0\rangle$.  The production of dual Fock state BECs in a double well potential was first proposed in \cite{Spekkens1999}.  This number squeezed state could be achieved by slowly applying a double well trapping potential to a condensate so that a phase transition occurs to the Mott insulator state and has been demonstrated experimentally in three-dimensional optical lattices \cite{Greiner2002}.  Initially the barriers between the three sites are high and, as in scheme 1, a two port beam splitter is applied to the first two modes of $|\psi\rangle_{B0}$.  The resulting output is sometimes referred to as a `bat' state since a plot of the amplitudes in the number basis resemble the ears of a bat.

Steps two to seven are identical to scheme 1 resulting in,
\begin{equation}
|\psi\rangle_{B4} = \frac{1}{\sqrt{2^N}(N/2)!} \left((\alpha_{-1}^\dag)^2 + e^{i2\phi}(\alpha_{1}^\dag)^2 \right)^{N/2}|0,0,0\rangle
\end{equation}
and
\begin{equation}
|\psi\rangle_{B7} = \frac{1}{2^N \left(N/2\right)!} \left( \left(a_0^{\dag} -ia_1^{\dag}\right)^2 + e^{i2\phi} \left(-ia_0^{\dag} +a_1^{\dag}\right)^2 \right)^{N/2} |0,0,0\rangle
\end{equation}
where a global phase has been ignored and $\phi$ is again given by $\phi = 2\sqrt{3}Jt_{\omega}\sin(\theta/3)$.  This is very similar to the scheme in \cite{Dunningham04}, where a bat state is used to measure a phase difference in a general Mach-Zehnder interferometer set-up, and in fact results in the same output.  The difference between the schemes is that ours has been adapted to measure a rotation around a ring of lattice sites rather than a general phase between two paths.

To determine $\phi$ we could count the number of atoms detected at each site and repeat.  However, this requires nearly perfect detector efficiencies \cite{Kim1999} and so is experimentally challenging.  Instead we use the read-out scheme described (in detail) in \cite{Dunningham04} to make our measurements.  

Essentially, after step seven the system is then left to evolve with the barriers high for $\tau= \pi/16V$ (note in the original paper $\tau=\pi/8U$ because $U=2V$ here).  The trapping potentials are then switched off and after some expansion time interference fringes are recorded.  These two steps are the read-out steps and are what we shall refer to as step eight.  The scheme is repeated many times and the visibility, $V$, of the fringes is calculated as in \cite{Dunningham04}.  From these visibility measurements we determine $\phi$ directly.  Note that the quantum Fisher information depends only on the final state of the system and not on the measurement procedure.

As such we use $|\psi\rangle_{B4}$ to determine the precision scaling of $\theta$ with $N$ and find $\Delta\theta \sim \sqrt{3}/(2Jt_{\omega}\cos(\theta/3)\sqrt{N(N/2+1)})$.  This means the uncertainty in $\omega$ is,
\begin{equation}
\Delta\omega \sim \left(\frac{h}{L^2 m}\right) \frac{\sqrt{3}}{2Jt_{\omega} \cos(\theta/3)\sqrt{N(N/2+1)}} \sim
 \left(\frac{h}{L^2 m}\right) \frac{\sqrt{3}}{\sqrt{2}Jt_{\omega}N},
\label{omegaBAT}
\end{equation}
where we have made the approximations $N\gg 1$ and $\theta/3 \ll 1$.  This has the same number scaling as the Heisenberg limit.

We now present a third scheme which offers a slight improvement in the precision scaling of our measurements.  We then consider the advantages and disadvantages of schemes 2 and 3 and discuss their experimental limitations.

%===================================================================================================================================
\section{Scheme 3: The NOON state}\label{cat}

This scheme is very similar to scheme 1, the only difference is that the two port 50:50 beam splitter (and its inverse) is replaced with a two port quantum beam splitter (and an inverse two port quantum beam splitter).  A two port quantum beam splitter (QBS) is defined as a device \cite{Dunningham} that outputs $(|N, 0\rangle+e^{i\xi}|0, N\rangle)/\sqrt{2}$ when $|N, 0\rangle$ is inputted.  The terms in the kets represent the number of particles in the two modes.  This output is called a NOON state.

As in scheme 1, the three potential barriers are initially high and $N$ atoms are inputted into site zero, $|\psi\rangle_{N0}=|N,0,0\rangle$.  The first step is to apply the QBS.  This is done using a scheme proposed in \cite{Dunningham01a}.  

The QBS begins with the application of a two port 50:50 beam splitter between sites zero and one, as previously described.  A $\pi/2$ phase is then applied to one of the two sites using an energy offset as above.  At this stage $V \gg J$ and the interactions are tuned such that their strength on one site is an integer multiple of the strength on the second site \footnote{This ensures that the required superposition is created independent of the total number of atoms, $N$}.  The system is left to evolve for $t=\pi/2V$ in this regime after which a second two port beam splitter is applied.  These steps output,
\begin{equation}
|\psi\rangle_{N1} \to \frac{1}{\sqrt{2N!}}\left((a_0^\dag)^N + e^{i\xi}(a_1^\dag)^N \right)|0,0,0\rangle .
\label{2site}
\end{equation}
where $\xi$ is some relative phase established by the splitting procedure.

Here we have a superposition of all $N$ atoms on site zero and all on site one.  At the equivalent stage in scheme 1 we had $N$ single particle superpositions (see equation \ref{psi1s1}).  It is this difference that is responsible for the improved precision. 

Steps two to six are identical to those in scheme 1 giving,
\begin{equation}
|\psi\rangle_{N4} = \frac{1}{\sqrt{2N!}} \left((\alpha_{-1}^\dag)^N + e^{i\xi}e^{iN\phi}(\alpha_{1}^\dag)^N  \right)|0,0,0\rangle .
\label{output}
\end{equation}
and
\begin{equation}
|\psi\rangle_{N6} = \frac{1}{\sqrt{2N!}} \left((a_0^\dag)^N + e^{i\xi}e^{iN\phi}(a_1^\dag)^N \right)|0,0,0\rangle .
\label{output}
\end{equation}

Finally we apply an inverse two port quantum beam splitter to the system.  This is achieved by sequentially undoing all the steps of the QBS.  So,\\
1. Apply an inverse two port beam splitter. \\
2. Raise the barriers and tune the interactions as before. \\
3. Leave the system to evolve for $t=\pi/2V$. \\
4. Apply a $-\pi/2$ phase to the same site as before. \\
5. Apply a second inverse two port beam splitter. \\
6. Raise the barriers.

This results in all $N$ atoms detected at site zero or all at site one with respective probabilities,
\begin{eqnarray}
P_0 = \cos^2\left(\frac{N\phi}{2} \right) \\ \nonumber
P_1 = \sin^2\left(\frac{N\phi}{2} \right) .
\end{eqnarray}

By repeating the scheme many times, each time recording the site on which all $N$ atoms are detected, $\phi$, and hence $\omega$, can be determined.  Using the quantum Fisher information and the Cramer-Rao lower bound the maximum resolution of scheme 3 is $\Delta \theta \sim \sqrt{3}/(2Jt_{\omega}\cos(\theta/3)N)$ meaning, for $\theta/3 \ll 1$,
\begin{equation}
\Delta \omega \sim \left(\frac{h}{L^2 m}\right)\frac{\sqrt{3}}{2Jt_{\omega}N}.
\label{omegaNOON}
\end{equation}  

We see the NOON state offers a slight improvement in resolution scaling over scheme 2.  It has the same number scaling, but the numerical factor is $\sqrt{2}$ better. However, this slight improvement in resolution may come at great experimental expense.  We now investigate the effect of experimental limitations on schemes 2 and 3.

%===================================================================================================================================

\section{Comparisons and practical limitations}\label{practical}
Both schemes 2 and 3 allow Heisenberg limited precision measurements of small rotations with the precision capabilities of scheme 3 being marginally favorable.  Our descriptions thus far, however, have only considered the idealized case and we have neglected important physical processes that may limit the experimental feasibility of the schemes.  We now reevaluate the schemes when these limitations are accounted for.

\subsection{Particle loss}
We need to account for the fact that atoms may be lost during the schemes.  It is well known that NOON states undergoing particle loss decohere quickly and so soon lose their Heisenberg limited sensitivity.  However, it has previously been shown that certain entangled Fock states, that allow sub-shot noise phase measurements, are much more robust to these losses \cite{Huver2008}.  Here we wish to see how resistant the bat state is to particle loss in comparison to the NOON state by determining the precisions afforded by both schemes in the presence of loss.

We investigate the effects of loss using a procedure described in \cite{Demkowicz09, Dorner09} which models loss from the middle section (between the two beam splitters) of an ordinary two path interferometer using fictitious beam splitters whose transmissivities, $\eta$, are directly related to the rate of loss.  The equivalent loss in our schemes is loss from the momentum modes during $t_{\omega}$. It was shown in \cite{Demkowicz09, Dorner09} that the time at which loss occurs during this interval is irrelevant.  Since losses are equally likely from both modes we consider equal loss rates, $\eta$, from each momentum mode.

To compare the effects of particle loss on our two schemes we determine $F_Q$ and $\Delta \phi$ for each scheme for different loss rates, or different $\eta$.  As shown in \cite{Demkowicz09, Dorner09} $F_Q$ varies with $\eta$ as,
\begin{equation}
F_Q= \sum_{l=0}^{N}F_Q\left[ \sum_{l_{\alpha_1}=0}^{l}p_{l_{\alpha_1},l-l_{\alpha_1}}|\xi_{l_{\alpha_1},l-l_{\alpha_1}}(\phi)\rangle \langle \xi_{l_{\alpha_1},l-l_{\alpha_1}}(\phi)| \right]
\end{equation}
where $l$ is the total number of atoms lost, $l_{\alpha_1}$ is the number lost from the $\alpha_1$ mode, $p_{l_{\alpha_1},l-l_{\alpha_1}}$ is the probability of each loss event and
\begin{equation}
|\xi_{l_{\alpha_1},l-l_{\alpha_1}}(\phi)\rangle = \frac{1}{\sqrt{p_{l_{\alpha_1},l-l_{\alpha_1}}}} \sum_{m=l_{\alpha_1}}^{N-(l-l_{\alpha_1})}\beta_me^{im\phi}\sqrt{B_{l_{\alpha_1},l-l_{\alpha_1}}^m} |m-l_{\alpha_1}, N-m-(l-l_{\alpha_1})\rangle.
\end{equation}
Here 
\begin{equation}
B_{l_{\alpha_1},l-l_{\alpha_1}}^m=\binom{m}{l_{\alpha_1}}\binom{N-m}{l-l_{\alpha_1}}\eta^N(\eta^{-1}-1)^l
\end{equation}
 and $\beta_m$ is
\begin{equation}
\beta_m=\frac{i^{N/2}\sqrt{m!}\sqrt{(N-m)!}}{2^{N/2}(m/2)!(N/2-m/2)!}\times \frac{1+(-1)^m}{2}
\end{equation}
for the bat state and
\begin{equation*}
\beta_m = \left\{
\begin{array}{rl}
1/\sqrt{2} & \text{for} \  m=0,N\\
0 & \text{for}  \ m\ne0,N 
\end{array} \right.
\end{equation*}
for the NOON state.  $F_Q$ is found numerically using 
\begin{equation}
F_Q = Tr[\rho(\phi)A^2]
\end{equation}
where $A$ is the symmetric logarithmic derivative which is defined by,
\begin{equation}
\frac{\partial \rho(\phi)}{\partial \phi} = \frac{1}{2}[A\rho(\phi) + \rho(\phi)A].
\end{equation}
It is given by,
\begin{equation}
(A)_{ij} = \frac{2}{\lambda_i + \lambda_j}\left[\frac{\partial\rho(\phi)}{\partial \phi}\right]_{ij}
\end{equation}
in the eigenbasis of $\rho(\phi)$, where $\lambda_{i,j}$ are the eigenvalues of $\rho(\phi)$.  Combining these equations gives \cite{Ping2007},
\begin{equation}
F_Q = \sum_{i,j}\frac{2}{\lambda_i+\lambda_j}\left|\left \langle \Psi_i \left| \frac{\partial \rho(\phi)}{\partial \phi} \right| \Psi_j \right \rangle \right|^2
\end{equation}
where $\Psi_{i,j}$ are the eigenvectors of $\rho(\phi)$.

\begin{figure}[h]
\centering
\includegraphics[scale=0.4]{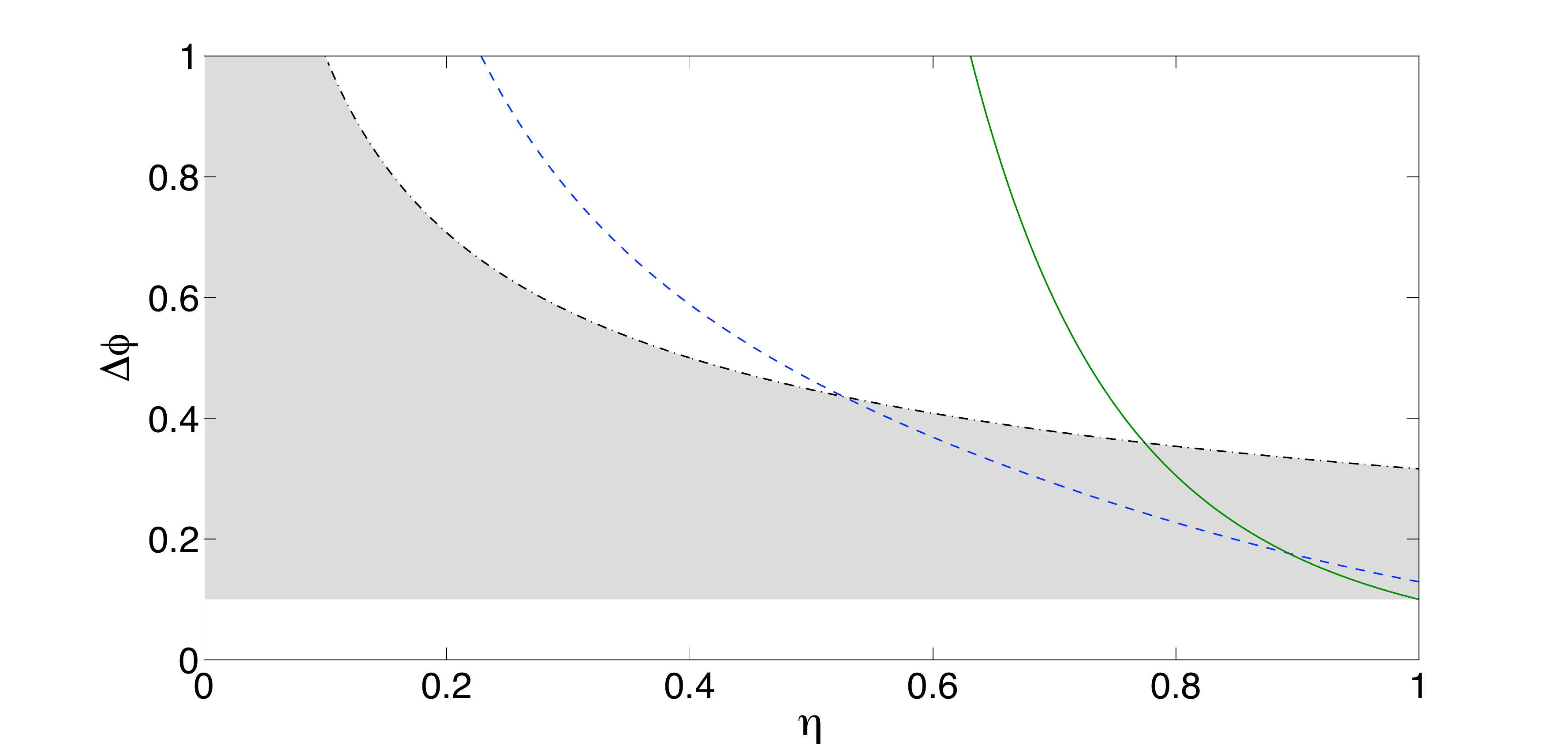}
\caption{The uncertainty of $\phi$ for different rates of loss, $\eta$, for $N=10$.  The blue dashed line shows $\Delta\phi$ for scheme 2 and the green solid line shows $\Delta\phi$ for scheme 3.  The upper bound of the shaded region is the precision afforded by scheme 1 (the classical precision limit - black dashed-dotted line) and the lower bound shows the Heisenberg limit.  Scheme 3 soon becomes less favorable than scheme 1, whilst scheme 2 is much more robust to losses.}
\label{BSloss}
\end{figure}

Figure \ref{BSloss} shows how $\Delta\phi$ varies with $\eta$ for the two schemes when $N=10$.  As expected the NOON state achieves the best precision when $\eta=1$, or when there are no losses.  However, as $\eta$ decreases the bat state soon becomes the favored scheme.  The lower bound of the shaded area is the Heisenberg limit and the upper bound is the precision achievable when an uncorrelated, or classical, input state is used, as in scheme 1.  We see the uncorrelated state soon outperforms the NOON state.  However, the bat state outperforms the uncorrelated input for approximately half the loss rates shown.  Since it is unlikely half the atoms would be lost in an experiment, the bat state, unlike the NOON state, appears to offer an experimentally feasible increase in precision over classical precision measurement experiments.  In the remainder of the paper we therefore assess the impact of experimental limitations on just scheme 2.

\subsection{Variations in N between experimental runs}
Our scheme requires many repetitions of the gyroscope procedure in order to build up interference fringes from which $\phi$ can be determined.  We have assumed thus far that each run involves exactly $N$ atoms.  However, in an experiment $N$ is likely to fluctuate between runs.  The effect of fluctuations of order $\sqrt{N}$ on the bat state input are discussed in \cite{Dunningham04}. In that case, an ordinary two-path linear interferometer is used but the same results apply here.  It is shown that while the interference fringe signal is degraded by these fluctuations, the approximate Heisenberg limited sensitivity of the scheme is not destroyed.  As expected, the larger $N$, the smaller the fluctuation effects, which is good since we would ideally work in the limit of large $N$ since this gives the best improvement in precision.

\subsection{Interactions}\label{Interactions}
So far we have considered only the idealized system setting $V=0$ (apart from in the detection step, step eight, where we require large interactions to minimize small coupling effects), and $J=0$ in the low coupling regime.  While interactions can be tuned to extremely small values using Feshbach resonances it is an unrealistic assumption to discount them altogether.  Likewise it is unrealistic to completely neglect coupling effects in the low coupling regime.  Here we consider the effect of non-zero interactions and non-zero coupling strengths in the low coupling regime on scheme 2.  To determine experimental orders of magnitude for $V$ and $J$ we use the approximations \cite{Scheel06}, 
\begin{equation}
V \approx \frac{2a_{s}V_{0}^{\frac{3}{4}}E_{R}^{\frac{1}{4}}}{\hbar(\sqrt{\lambda D})} 
\end{equation}
\begin{equation}
J \approx \frac{E_R}{2\hbar}\exp\left(-\left(\frac{\pi^2}{4}\right)\sqrt{\frac{V_0}{E_R}}\right)\left(\sqrt{\frac{V_0}{E_R}}+\left(\sqrt{\frac{V_0}{E_R}}\right)^3\right)
\end{equation}
where $a_{s}$ is the scattering length, $V_{0}$ the barrier height, $E_{R}$ the recoil energy, $\lambda$ the wavelength of the lattice light and $D$ the transverse width of the lattice sites.  Feshbach resonances can tune $a_{s}$ to values smaller than the Bohr radius for some BECs \cite{Chin08} and in the high coupling regime barrier heights of order $V_{0}=2E_{R}$ have been demonstrated \cite{Jona03}.  Using light of wavelength $\lambda=D=10\mu$m and ${}^{87}$Rb atoms, interactions can therefore be tuned to $V \sim 10^{-3}$Hz and $J \sim 10$Hz in this regime.  While in the low coupling regime, where $V_{0}=35E_{R}$ \cite{Greiner2002a}, we find $V \sim 10^{-2}$Hz and $J \sim 10^{-2}$Hz.  Note that in the detection process the system evolves for $t=\pi/16V$ with high potential barriers.  Here we require large interactions to minimize small coupling effects.  Taking $a_s=9000a_0$ \cite{Cornish2000} gives $V \sim 100$Hz.  Using these values we assess the impact of non-zero interactions and coupling strengths.

\begin{figure}[h]
\centering
\includegraphics[scale=0.4]{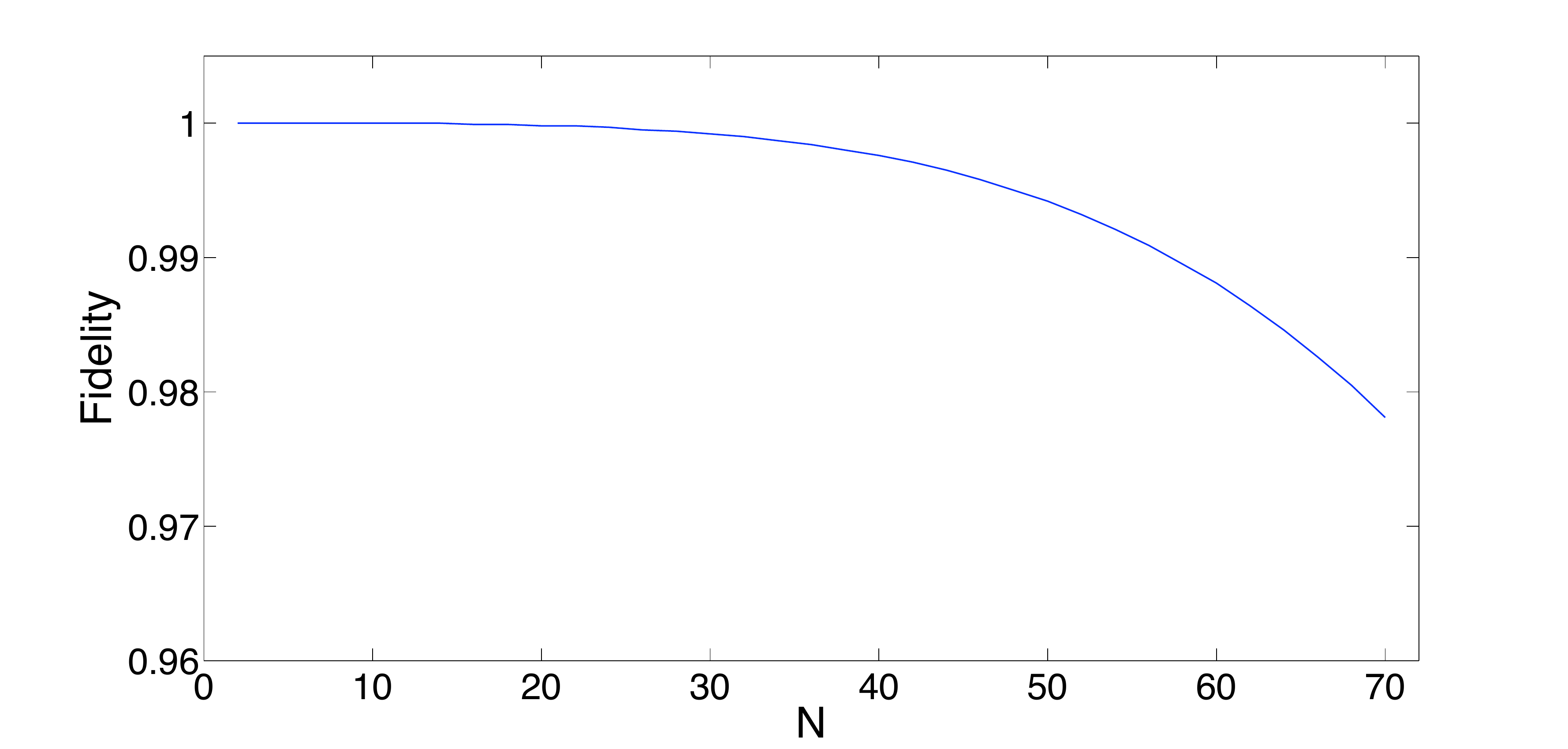}
\caption{The fidelity between the output of scheme 2 in the idealized case (where $V=0$, and $J=0$ in the low coupling regime) with the output in the non-idealized case (where $V \sim 10^{-3}$ in the high coupling regime, $V \sim 10^{-2}$ in the low coupling regime and $J \sim 10^{-2}$ in the low coupling regime) for different numbers of input atoms. Here $\theta=\pi/100$ and $t_{\omega}=1$s.}
\label{interactionN}
\end{figure}

As $N$ increases the occupation number per site increases and as such the effect of non-zero interactions become more pronounced.  We would therefore like to determine the maximum number of atoms our system can tolerate before these effects become too destructive.  To do this we measure the fidelity between the output of the gyroscope in the idealized case in which $V=0$ (except in step eight) and $J=0$ in the low coupling regime with the same output when $V \sim 10^{-3}$ in the high coupling regime, $V \sim 10^{-2}$ in the low coupling regime and $J \sim 10^{-2}$ in the low coupling regime.  The maximum $N$ that can be tolerated is taken to be the first $N$ for which this fidelity falls below 0.99.  In this simulation we have taken $t_{\omega} = 1$s and $\theta=\pi/100$.  Figure \ref{interactionN} shows how the fidelities decrease as $N$ increases.  We see that by our definition the maximum number of atoms the system can tolerate is approximately 60.  Squeezed states with larger numbers of atoms have been demonstrated experimentally \cite{Esteve2008} and as such interactions are likely to be one of the main limiting factors of this scheme.

\subsection{Metastability}
Our gyroscope scheme relies on the two counterpropagating  superfuid states being stable at least over the 
timescale of the measurement that we want to make. There has been a lot of discussion and analysis of what conditions need to be met so that a superfluid current persists (or is metastable) in BECs. In a toroidal trap, for example, the condition for metastability is that the mean interaction energy per particle is greater than the single particle quantization energy $h^2/mL^2$ \cite{Leggett2001a}. This can be shown to be equivalent to the condition that the superfluid velocity is less than the velocity of sound, which is just the Landau criterion for the metastability of superfluid flow.  Persistent superfluid flows have recently been experimentally observed for BECs in toroidal traps \cite{Helmerson2007a}.

There has also been an analysis of the metastability requirements for a system that is directly relevant to our scheme, i.e. a superfluid Bose gas trapped in a ring optical lattice \cite{Kolovsky2006a}.  In this work, it was shown that any disorder due to energy offsets, $\epsilon$, on the lattice sites in the ring can cause dissipation by providing a coupling pathway between states with opposite quasi-momentum. If, however, there are interactions between the atoms, the energy levels are shifted and this mismatch means that the different supercurrent states become effectively decoupled and so the supercurrent should persist. Kolovsky showed that for states with low values of quasimomentum (such as the $\pm$1 unit of angular momentum states that we consider) the minimum strength of the interactions must be $V_{\rm min} \approx 8\epsilon/N$. Taking parameters from our scheme, in the strong tunneling region, we have $J \sim10$Hz and $V \sim 10^{-3}$Hz. This means that we require $\epsilon < N V/8 \approx 10^{-2}$Hz, for $N\approx 60$. By stabilizing the energy offsets to this level (or increasing the interactions) persistent  supercurrents should be able to be achieved. 

\subsection{Comparison with other schemes}
At this point we note that our precision analysis is for the case of a single shot, i.e. $N$ atoms are loaded into the lattice and a single measurement is made over time $t_{\omega}$ corresponding to a particle flux of $N/t_{\omega}$. In reality, the results of many runs will be combined to give a measurement of the rotation. Suppose we repeat the measurement $n$ times to give a total integration time of $\tau = nt_{\omega}$\footnote{Note that there are still $N$ atoms on each run, which lasts for time $t_{\omega}$, so the particle flux is the same as for the case of a single run.}. In this case, we get
\begin{equation}
\Delta\omega \approx 
 \left(\frac{h}{L^2 m}\right) \frac{\sqrt{3}}{\sqrt{2n}Jt_{\omega}N} = 
  \left(\frac{h}{L^2 m}\right) \frac{\sqrt{3}}{\sqrt{2t_{\omega} \tau}JN} = \frac{S}{\sqrt{\tau}},
\end{equation}
where the short-time sensitivity is given by,
\begin{equation}
S =  \left(\frac{h}{L^2 m}\right) \frac{\sqrt{3}}{\sqrt{2t_{\omega}}JN}.
\end{equation}
Substituting in approximate values for our setup in the strong coupling regime (i.e. $J\approx 10$Hz, $N\approx 60$, $t_{\omega}\approx 1$s and $L=2\pi \times 20 \mu$m \cite{Henderson2009}), we get $S\approx 10^{-3}$ rads$^{-1}/\sqrt{\rm{Hz}}$.
This compares unfavorably with other atom interferometry schemes which can achieve sensitivities better than $10^{-8}$ rads$^{-1}/\sqrt{\rm{Hz}}$ \cite{Gustavson1997}. These other schemes achieve improved sensitivities by having much larger particle fluxes (e.g. $6\times 10^{8}$ atoms/s) and much larger areas enclosed by their interferometer paths (e.g. 22 mm${}^2$) \cite{Gustavson1997}.

The scheme presented here is therefore unlikely to challenge the overall precision offered by other techniques, except perhaps in specialised cases where the number of atoms available is restricted to a small number or the area of the interferometer must be very small. The main interest of this scheme, however, is that it
proposes a means of creating macroscopic superpositions of persistent superfluid flows and that these could show evidence of Heisenberg scaling of measurement precision. This in itself would be of fundamental interest. In order to improve its short-term sensitivity, however, it is likely to be difficult to create entangled states with very large numbers of particles, so a different configuration would need to be used to greatly enhance the enclosed area of the interferometer.  

Alternatively nonlinear couplings of the atoms and the rotation could potentially be used to achieve precision scalings of $N^{-3/2}$ using unentangled atoms \cite{Boixo2008, Boixo2009a}.  The reliance of such a scheme on unentangled atoms would allow for much larger atom numbers than in our scheme and as such seems to offer a possible way of improving upon current gyroscope precisions.  However, the challenge of this scheme would be coupling the angular momentum and the atomic scattering lengths.

\section{Conclusion}
We have presented three schemes to measure small rotations applied to a ring of lattice sites by creating superpositions of ultra-cold atoms flowing in opposite directions around the ring.  Two of these schemes are capable of Heisenberg limited precision measurements where the precision scales as $1/N$.  The two schemes use different entangled states.  While the scheme that used a NOON state gave slightly better precision in the idealized case, after consideration of experimental limitations it was shown that the bat state is likely to be the preferred candidate largely due to its robustness to particle loss.  Importantly the bat state outperformed the case of unentangled particles for modest loss rates.  The effects of non-zero interactions were shown to limit the preferred scheme to approximately 60 atoms and as such this scheme is not capable of outperforming the precision of existing atomic gyroscopes at present.  However, the interesting result is the Heisenberg scaling of the precision.  All the steps in this scheme should be within reach of current technologies which is promising for its experimental implementation.

This work was supported by the United Kingdom EPSRC through an Advanced Research Fellowship GR/S99297/01, an RCUK Fellowship, and the EuroQUASAR programme EP/G028427/1.

%=======================================================================================================================

%=======================================================================================================================

\end{document}